# Challenges and Lessons Learned from fabrication, testing and analysis of eight MQXFA Low Beta Quadrupole magnets for HL-LHC

G. Ambrosio, K. Amm, M. Anerella, G. Apollinari, G. Arnau Izquierdo, M. Baldini, A. Ballarino, C. Barth, A. Ben Yahia, J. Blowers, P. Borges De Sousa, R. Bossert, B. Bulat, R. Carcagno, D. W. Cheng, G. Chlachidze, L. Cooley, M. Crouvizier, A. Devred, J. DiMarco, S. Feher, P. Ferracin, J. Ferradas Troitino, L. Garcia Fajardo, S. Gourlay, H. M. Hocker, S. Izquierdo Bermudez, P. Joshi, S. Krave, E. M. Lee, J. Levitan, V. Lombardo, J. Lu, M. Marchevsky, V. Marinozzi, A. Moros, J. Muratore, M. Naus, F. Nobrega, T. Page, I. Pong, J.C. Perez, S. Prestemon, K. L. Ray, G. Sabbi, J. Schmalzle, J. Seyl, S. Sgobba, S. Stoynev, T. Strauss, E. Todesco, D. Turrioni, G. Vallone, R. Van Weelderen, P. Wanderer, X. Wang, M. Yu

***Abstract*—By the end of October 2022, the US HL-LHC Accelerator Upgrade Project (AUP) had completed fabrication of ten MQXFA magnets and tested eight of them. The MQXFA magnets are the low beta quadrupole magnets to be used in the Q1 and Q3 Inner Triplet elements of the High Luminosity LHC. This AUP effort is shared by BNL, Fermilab, and LBNL, with strand verification tests at NHMFL.**

**An important step of the AUP QA plan is the testing of MQXFA magnets in a vertical cryostat at BNL. The acceptance criteria that could be tested at BNL were all met by the first four production magnets (MQXFA03-MQXFA06). Subsequently, two magnets (MQXFA07 and MQXFA08) did not meet some criteria and were disassembled. Lessons learned during the disassembly of MQXFA07 caused a revision to the assembly specifications that were used for MQXFA10 and subsequent magnets.**

**In this paper, we present a summary of: 1) the fabrication and test data of all the MQXFA magnets; 2) the analysis of MQXFA07/A08 test results with characterization of the limiting mechanism; 3) the outcome of the investigation, including the lessons learned during MQXFA07 disassembly; and 4) the finite element analysis correlating observations with test performance.**

This work was supported by the U.S. Department of Energy, Office of Science, Office of High Energy Physics, through the US LHC Accelerator Upgrade Project (AUP), and by the High Luminosity LHC project at CERN.

G. Ambrosio, G. Apollinari, M. Baldini, J. Blowers, R. Bossert, R. Carcagno, G. Chlachidze, J. DiMarco, S. Feher, S. Krave, V. Lombardo, V. Marinozzi, F. Nobrega, T. Page, J. Seyl, S. Stoynev, T. Strauss, D. Turrioni, M.Yu are with Fermi National Accelerator Laboratory, Batavia, IL 60510 USA, (e-mail: giorgioa@fnal.gov).

K. Amm, M. Anerella, A. Ben Yahia, H. M. Hocker, P. Joshi, J. Muratore, J. Schmalzle, P. Wanderer, are with Brookhaven National Laboratory, Upton, NY 11973 USA.

D. W. Cheng, P. Ferracin, L. Garcia Fajardo, E. M. Lee, M. Marchevsky, M. Naus, I. Pong, S. Prestemon, K. Ray, G. Sabbi, G. Vallone, X. Wang are with Lawrence Berkeley National Laboratory, Berkeley, CA 94720 USA.

L. D. Cooley, J. Levitan, J. Lu are with or affiliated to the Applied Superconductivity Center, National High Magnetic Laboratory, Tallahassee, FL 32310, USA and with FSU.

G. Arnau Izquierdo, A. Ballarino, C. Barth, B. Bulat, M. Crouvizier, A. Devred, J. Ferradas Troitino, A. Moros, S. Izquierdo Bermudez, J.C. Perez, E. Ravaioli, S. Sgobba, P. Tavares Coutinho Borges De Sousa, E. Todesco, R. Van Weelderen are with CERN, Geneve, Switzerland.

## I. Introduction

The HL-LHC project [1], aiming at 3000 $fb^{-1}$, is in full swing in all participating countries and institutions. The cornerstones of this project are the low-beta quadrupole magnets (MQXF) [2] with an unprecedentedly large aperture (150 mm) and gradient (132.6 T/m). MQXF are the first magnets to use $Nb_3Sn$ conductor in a particle accelerator, paving the way for the use of this material in future high-energy colliders. The main challenges facing these magnets include: 1) the electromagnetic forces they are subject to, which in the straight section and in the ends (1.15 MN) are four and six times higher, respectively, than in the LHC low-beta magnets, and 2) the stored energy per unit length (1.2 MJ/m), which is more than double the energy per unit length of the LHC main dipole magnets. The solutions to these design challenges have been presented elsewhere [3] and [4]. This paper focuses on the lessons learned during the fabrication and test of eight MQXFA magnets [5] that will be used in the Q1 and Q3 Inner Triplet (IT) elements of the HL-LHC. These magnets are fabricated in the US by the HL-LHC Accelerator Upgrade Project (AUP) [6]. CERN is fabricating the MQXFB magnets [7], with an almost identical cross section, for the Q2A/B IT elements. The lessons learned during the MQXFA prototyping phase have been presented in [8].

## II. Fabrication Process and Status

The MQXFA fabrication process includes activities performed at Brookhaven National Laboratory (BNL), Fermi National Accelerator Laboratory (FNAL), Lawrence Berkeley National Laboratory (LBNL) and the National High Magnetic Laboratory (NHMFL). The strand used in the MQXFA magnet is RRP 108/127 by Bruker-OST [9], [10]. Strand QC verification is done at NHMFL. Cables are fabricated by LBNL, and together with coil parts (procured by FNAL) are sent to either BNL or FNAL for coil fabrication. 50% of the coils are fabricated at BNL and 50% at FNAL with identical design and procedures so that they are fully replaceable. The coils are sent to LBNL, where structures are procured and the magnet assembled. The magnets are shipped to BNL for a vertical test, and subsequently to FNAL if they meet their acceptance criteria [11]. At FNAL coldmasses are assembled and inserted into cryostats provided by CERN. The cryoassemblies are tested at FNAL and shipped to CERN if they meet all of the acceptance criteria [12].

By the end of October 2022, the coil fabrication status was 76% complete, nine magnets (MQXFA03-11) had been assembled, eight magnets (MQXFA03-08 and 10-11) tested, three magnets (MQXFA07-09) disassembled, and one magnet reassembled after replacing a coil (MQXFA08b). During preshipment inspection of MQXFA09 it was found that a polyimide layer between coil midplanes had folded during coil-pack assembly, creating a small step. For this reason, MQXFA09 was disassembled for coil inspection and analysis.

## III. Vertical Test Results

The MQXFA quadrupoles are tested individually at the vertical magnet test facility of the Superconducting Magnet Division at BNL in superfluid He at 1.9 K and 1 bar. This test aims at verifying that each magnet meets the acceptance criteria [11] before it is used in an LMQXFA cold mass. The magnets are tested in a 6.1 m test Dewar fitted with a warm bore that allows the use of a quench antenna or rotating coils during testing. The magnets are powered by a 30-kA power supply (two 15-kA power supplies in parallel) equipped with an energy extraction circuit using IGBT switches [13]. During testing, quench protection is provided by an energy extraction system set at 37.5 mΩ with a 10-ms delay, quench protection heaters (QPH) at 600 V and 12.4 mF, and a Coupling Loss Induced Quench (CLIQ) system [14] set at 500 V and 40 mF. Two digital FPGA-based quench detectors are used to monitor the magnet half-voltage difference, the total magnet voltage, and the voltage of the various splices and of the superconducting leads. The MQXFA magnets are instrumented with voltage taps for quench detection and localization, strain gauges installed on coil poles, shells, and axial rods [5]. The test facility uses a quench antenna array with 50-mm longitudinal detection resolution [15], rotating coils for magnetic field measurements, and sensors for measurement of temperature, pressure, and liquid He level.

The main features of the test plan for every magnet are 1) cooldown and warmup while maintaining a maximum end-to-end gradient of 50 K; 2) training at 1.9 K with 20 A/s ramp rate to acceptance current (nominal current + 300 A = 16530 A); 3) holding tests at the acceptance current; 4) ramp to nominal current at 30 A/s and ramp down at 100 A/s; 5) ramp to nominal current at 4.5 K; 6) splices and magnetic measurements; 7) verification of the training memory after a thermal cycle; and 8) verification of all electrical requirements at the end of the test. An example of the detailed procedures used for each magnet test can be found in [16].

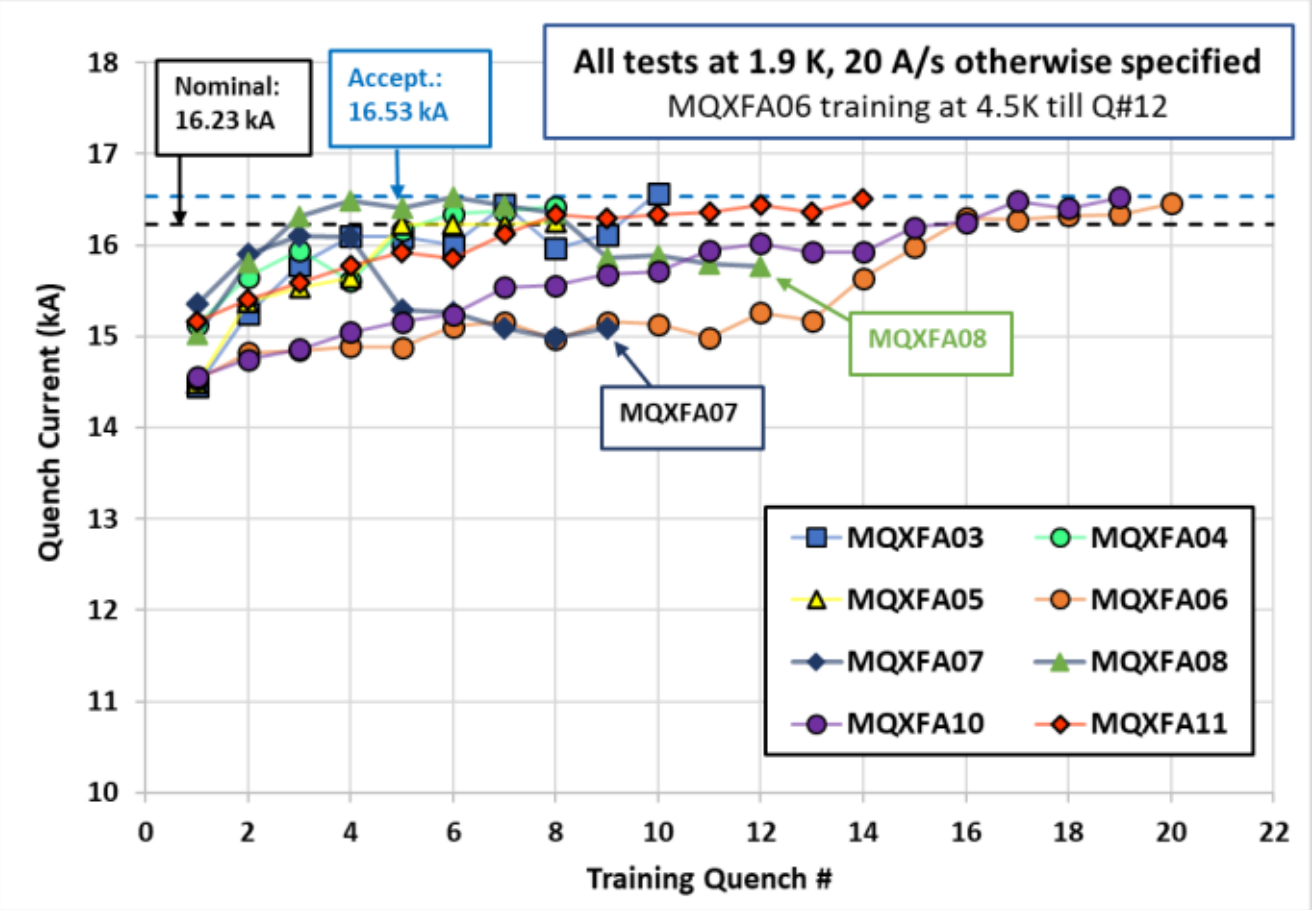

Fig. 1. Quench history of all the MQXFA magnets during vertical tests at BNL (only training quenches are shown). MQXFA06 was trained at 4.5 K until quench #12. MQXFA07 and MQXFA08 training can be seen in Figs. 2 - 3.

Eight MQXFA magnets were tested in the vertical test facility. Fig. 1 shows the training part of their tests. Six of the eight magnets, MQXFA03-06 and MQXFA10-11, reached and held acceptance current, while MQXFA07 and 08 showed limited performance. MQXFA05 went through an additional endurance test; it initially reached acceptance current after eight quenches and demonstrated good training memory after two thermal cycles. The endurance test included 42 induced quenches at nominal current (16.23 kA) performed over two additional thermal cycles. At the end of the endurance test, the magnet reached nominal current at 4.5K and again achieved acceptance current at 1.9 K.

The MQXFA07 quench history plot is shown in Fig 2. MQXFA07 started training at 15.3 kA and reached 16.1 kA after three quenches. Subsequently, during quench #5, it showed a detraining of 800 A. The following quenches at 1.9 K and with a 20 A/s ramp rate eventually settled at a quench current in the range 15–15.1 kA, showing a drop of about 1 kA with respect to the quench current reached in quenches #3 and #4.

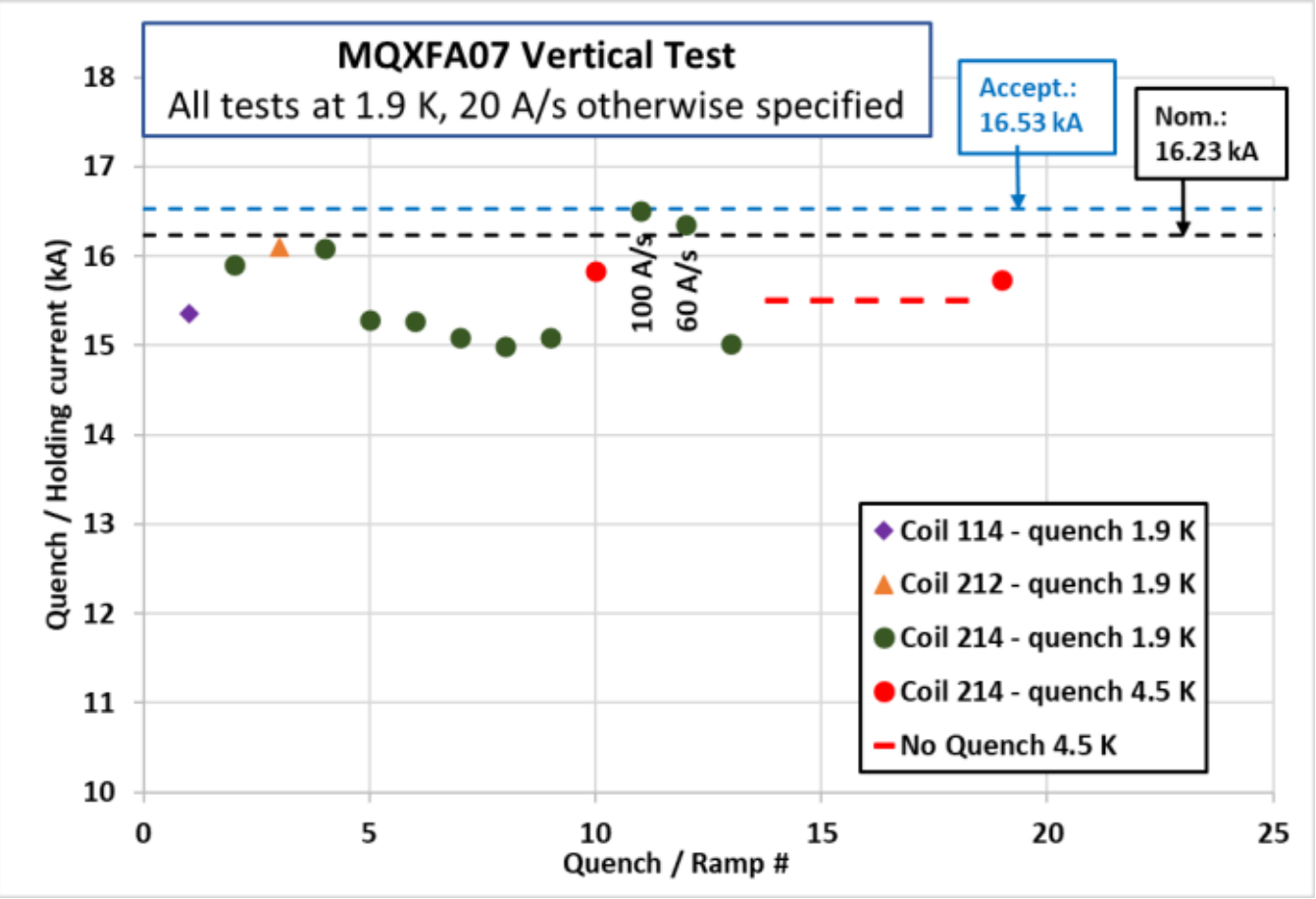

Fig. 2 MQXFA07 quench history plot.

All limiting quenches started in a single coil (Coil 214 in quadrant 3), in the same voltage-tap segment (A3-A4), and about 21 ms before the quench detection trigger. The quench antenna array showed that all limiting quenches in coil 214 at 1.9 K and 20 A/s took place in the same longitudinal position along the coil, at around 2050 mm from the magnet's center. This location corresponds to the beginning of the lead end of the coil. Fig. 2 shows reverse temperature dependence with the quench current at 4.5 K about 800 A higher than at 1.9 K at the standard 20 A/s ramp rate, and reverse ramp-rate dependence with quench currents 1486 A and 1332 A higher at 100 and 60 A/s, respectively, than the last quench at 20 A/s.

The MQXFA08 quench history plot is shown in Fig 3. The magnet started training at 15 kA and reached acceptance current (16.53 kA) after five quenches. During the hold at 16.53 kA, a power supply shutoff triggered the quench detection, inducing a quench in the magnet. Subsequently, the magnet exhibited detraining, with the quench current settling at around 15.7 kA after three quenches. All limiting quenches occurred in coil 213 in quadrant 3. While the intermediate voltage taps were not available initially, when they were reconnected after the second thermal cycle, they revealed that the limiting quenches were occurring in the same voltage-tap segment as in MQXFA07 (A3-A4) when ramping at the typical 20 A/s at 1.9 K. These quenches started about 16 ms before the quench detection trigger, and the quench antenna data showed that they all occurred in the lead end at about 2100 mm from the magnet's center. Although initially the quenches at 4.5 K were occurring at a lower current (14.5 kA), it was found that the splice between the magnet's negative lead and the test facility had resistance equal to 42 nΩ. Once the splice was repaired, during the third test cycle, MQXFA08 showed reverse temperature dependence with quench currents in the range of 16-16.4 kA at 4.5 K. At 1.9 K the magnet showed reverse ramp-rate dependence, with the highest quench current achieved at 60 A/s being 400 A higher than that at 20 A/s, and over 3 kA higher than the current reached at 5 A/s. At 4.5 K there was no clear trend in the quench currents in the rate range investigated during the test (5-100 A/s).

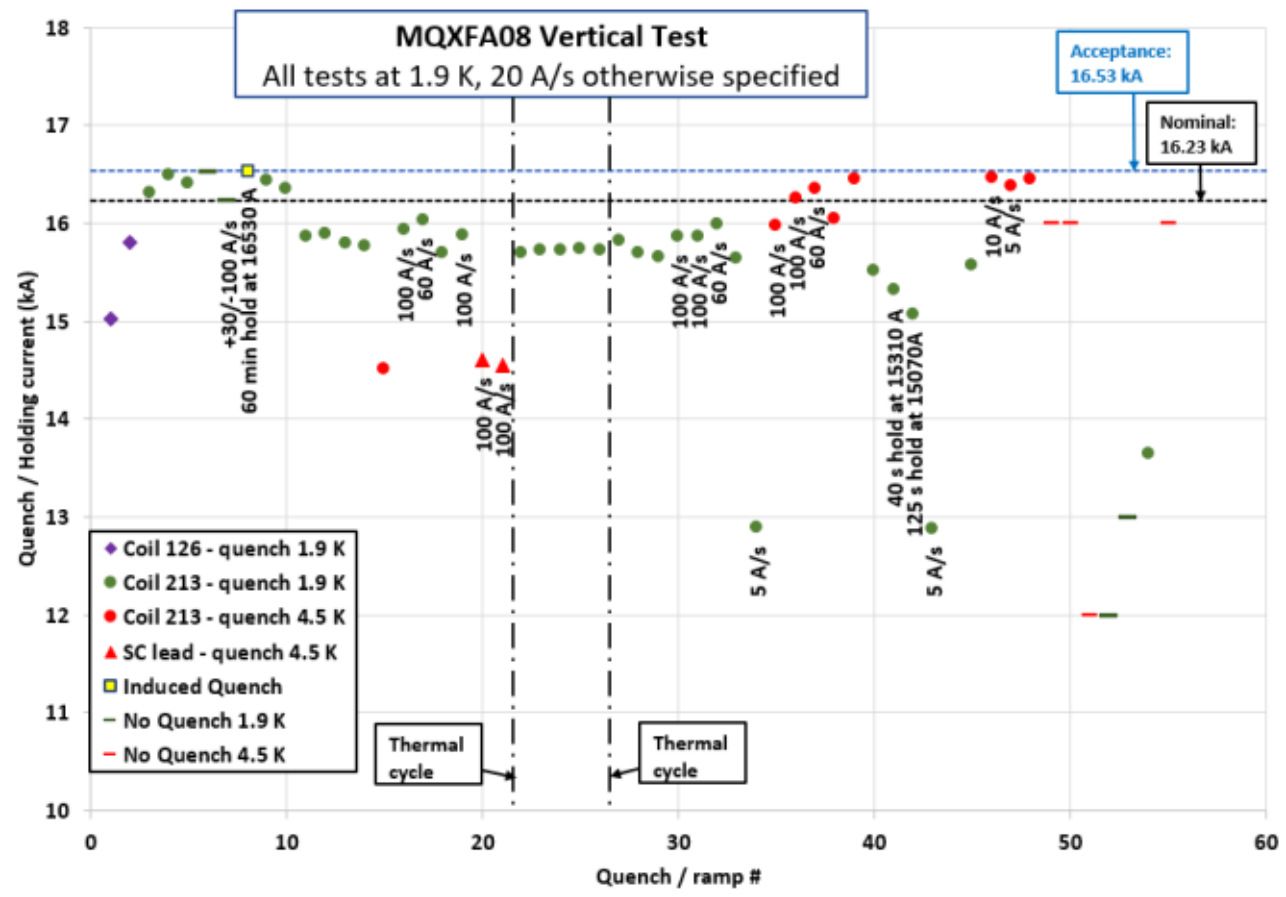

Fig. 3. MQXFA08 quench history plot.

The reverse temperature and ramp-rate dependences observed in MQXFA07 and MQXFA08 are clear signatures of the mechanism understood to be the cause of the limited performance: self-field instability [17], [18] enhanced by conductor damage. This is caused by conductor damage that pushes more current into a few strands (or in the not-damaged sections of a few strands), triggering the self-field instability in those strands. This mechanism was found in a LARP Long Quadrupole [19], although in that magnet the damage was likely in a low-field area, therefore triggering thermo-magnetic instability with its typical flux jumps. The coil autopsy presented in next section showed that the conductor was damaged where the MQXFA magnetic field is between 6 and 9 T at 15.3 kA. Measurements performed at CERN on leftovers of the strands used in the MQXFA07/A08 limiting coils showed no instability in this field range [20]. Therefore, we are assuming that a combination of uneven current distribution and partially damaged strands can explain the observed behavior.

An MQXF short model (MQXFS3a) [21] had a very similar quench history, with reverse temperature dependence, reverse ramp-rate dependences, quenches around 15 kA at 1.9 K and 20 A/s, and a quench start location in the pole block of the lead end of the limiting coil. The following analysis may also explain the behavior of MQXFS03.

## IV. MQXFA07/08 DISASSEMBLY

A thorough analysis of coil fabrication and QC data found no anomalies in coils 213 and 214 apart from the COVID lockdown impact reported in section VII.

After the test at BNL and the shipment to LBNL, both the MQXFA07 (A07) and MQXFA08 (A08) magnets were inspected, unloaded and disassembled. Several measurements were taken before, during, and after the unloading, with the goal of looking for any possible anomaly. The first surprising observation that emerged after removing the coil-pack subassembly from the shell-yoke sub-assembly (Fig. 4) was related to the coil-pack horizontal dimensions. As can be seen in Fig. 5, the A07 coil-pack exhibited a trapezoidal shape in the post ("p") test data (red markers) with the top (T) horizontal dimension being larger than the bottom (B).

This shape was also present during the first assembly procedure (the one preceding the test), as shown by the blue markers. Surprisingly, the difference between the top and bottom dimensions of the coil-pack was not corrected by the loading operation. In other words, the asymmetry remained despite the fact that after the bladder operation the coil-pack is locked by identical loading keys inside the square cavity of the shell-yoke sub-assembly.

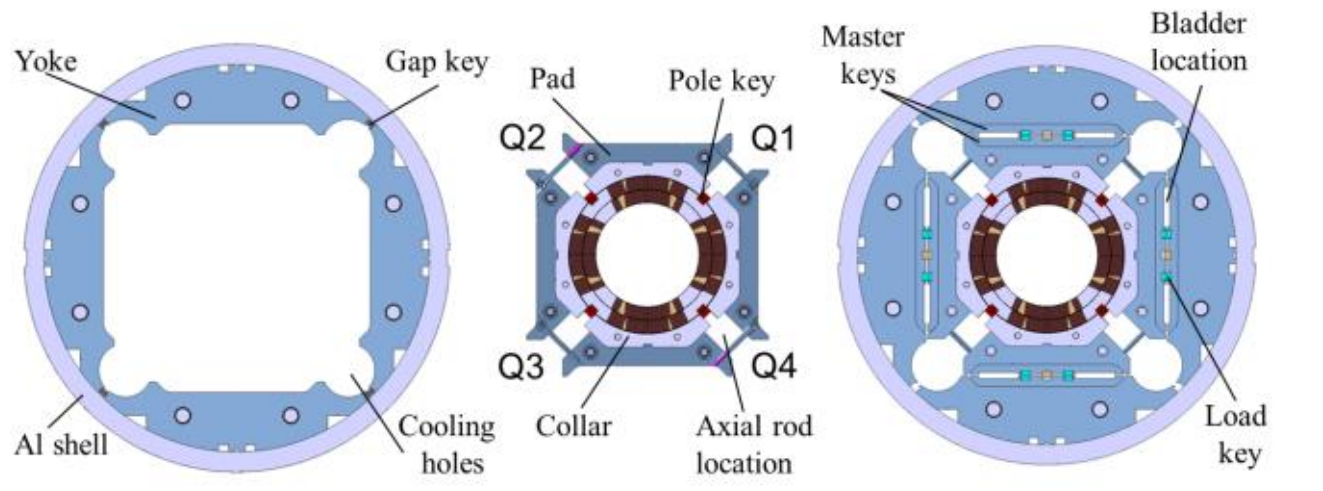

Fig. 4 Magnet cross-section (right), coil-pack sub-assembly (center), shell-yoke sub-assembly (left). Q1 to Q4 are coil quadrants as seen from the magnet's lead end.

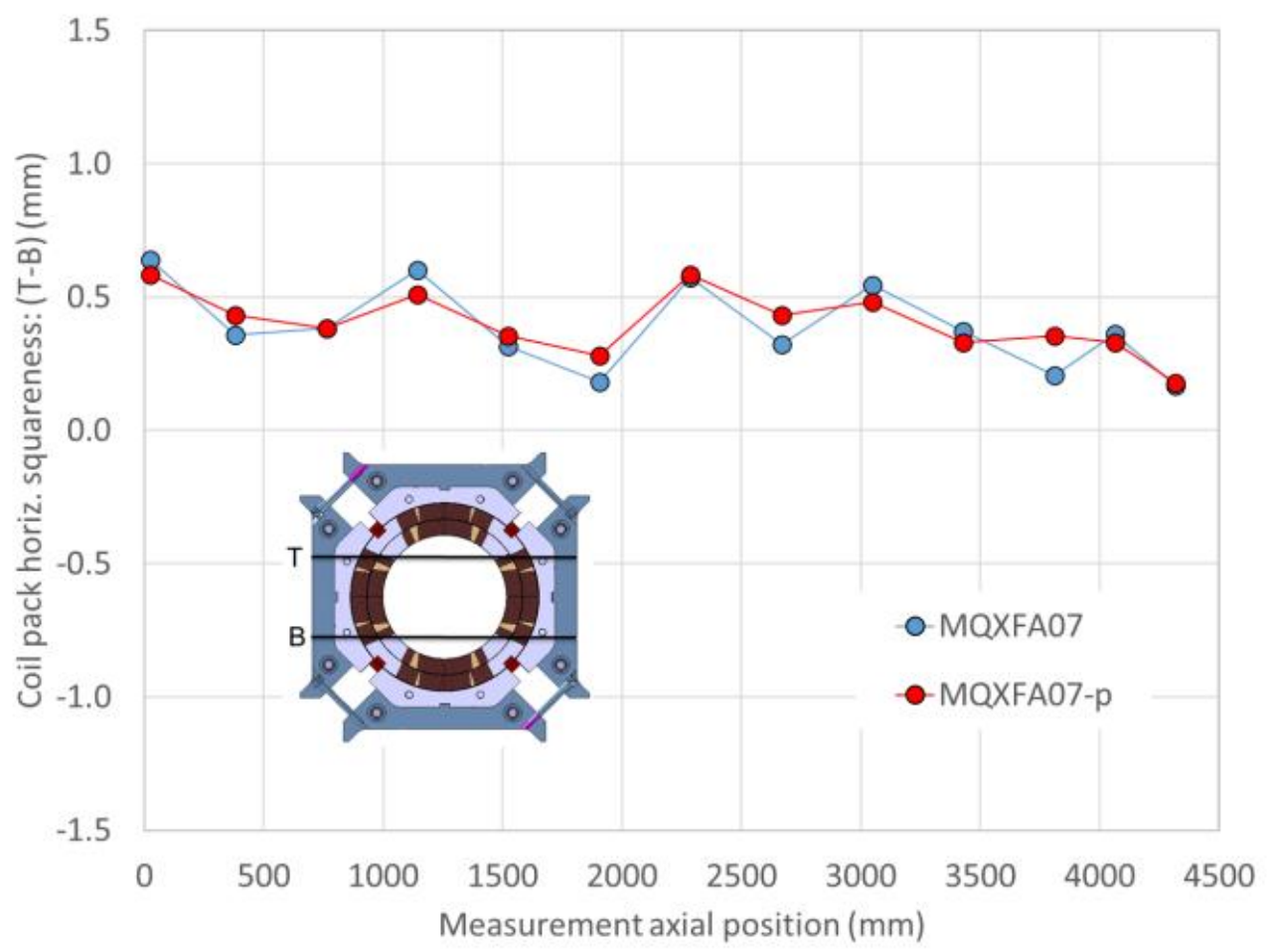

Fig. 5. Measurement of MQXFA07 coil-pack horizontal squareness before the test (A07) and after the test (A07-p). The horizontal squareness is the difference between the top (T) and the bottom (B) dimensions.

Consistently with the trapezoidal shape of the coil pack, a non-uniformity in the distribution of the pole-key gaps was observed after the vertical test. The pole-key gaps are small gaps between each pole key and the collars. Fig. 6 shows measurements of these gaps in each quadrant as a function of the axial position in MQXFA07 after the test. The specifications for the pole gaps used during the assembly of A07 and A08, indicated by the dashed lines, stated that the average pole-key gap (per key side) among the four coils on each longitudinal location shall be +0.200 ±0.050 mm. It can be seen that the average among the four quadrants was on spec, but a significant variation among quadrants was present. In particular, in quadrant 3 (Q3) a null gap was observed after the test. A similar very low pole-key gap was observed both before and after the test in A08 as well, again indicating that the pre-load operation did not correct this asymmetry.

After the coil pack disassembly, coils, ground isolation layers, and pole keys were inspected. In Q3, deep imprints in the polyimide insulation between the collars and pole keys, indicating collar lamination lines and pole key G11 grain, were observed (Fig. 7, top). In addition, the G11 pole keys in Q3 also showed high pressure imprints of collar lamination gaps (Fig. 7, bottom). These imprints, all indicative of higher pressure, were not seen in the other quadrants.

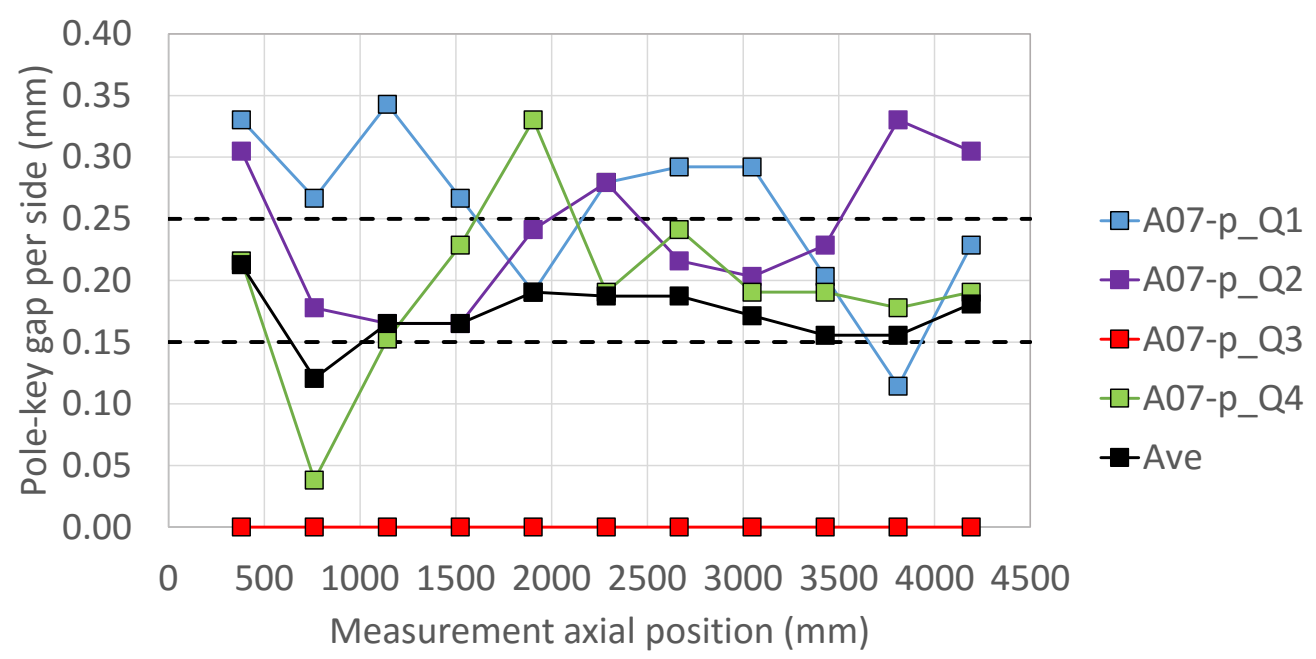

Fig. 6. Pole key gap measurements in each quadrant as a function of the axial position after the test (A07p). The dashed lines indicate the spec values of the average among the four quadrants on each longitudinal location.

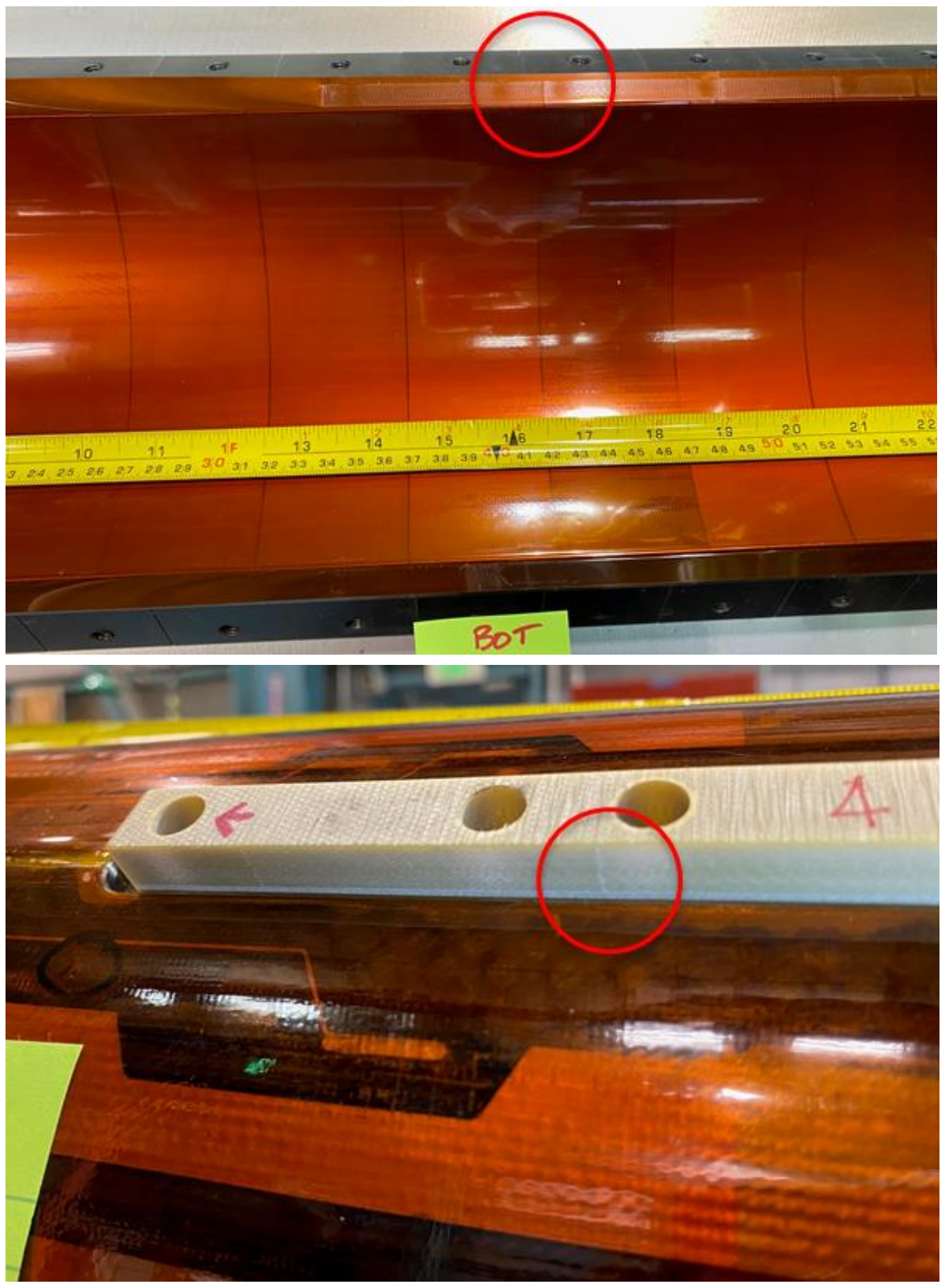

Fig. 7. Red circles: high pressure "imprints" of collar lamination on the polyimide insulation between the collars and pole keys (top) and on the pole keys of Q3 (bottom).

The visual inspection of coil 214 (the A07 limiting coil) found large bubbles (delamination of the insulation on the coil's inner surface) in both ends on the inner layer at the coil tips. These bubbles are typical in MQXF coils after testing. However, a closer look at both ends showed that there is a small bubble/delamination at each transition between wedge and end-spacer (see yellow circles in Fig. 8, left). In addition, a dye-penetrant test performed at CERN [22] to check for bonding among coil components showed horizontal cracks starting from the wedge to the end-spacer transitions (see red circles in Fig. 8, right).

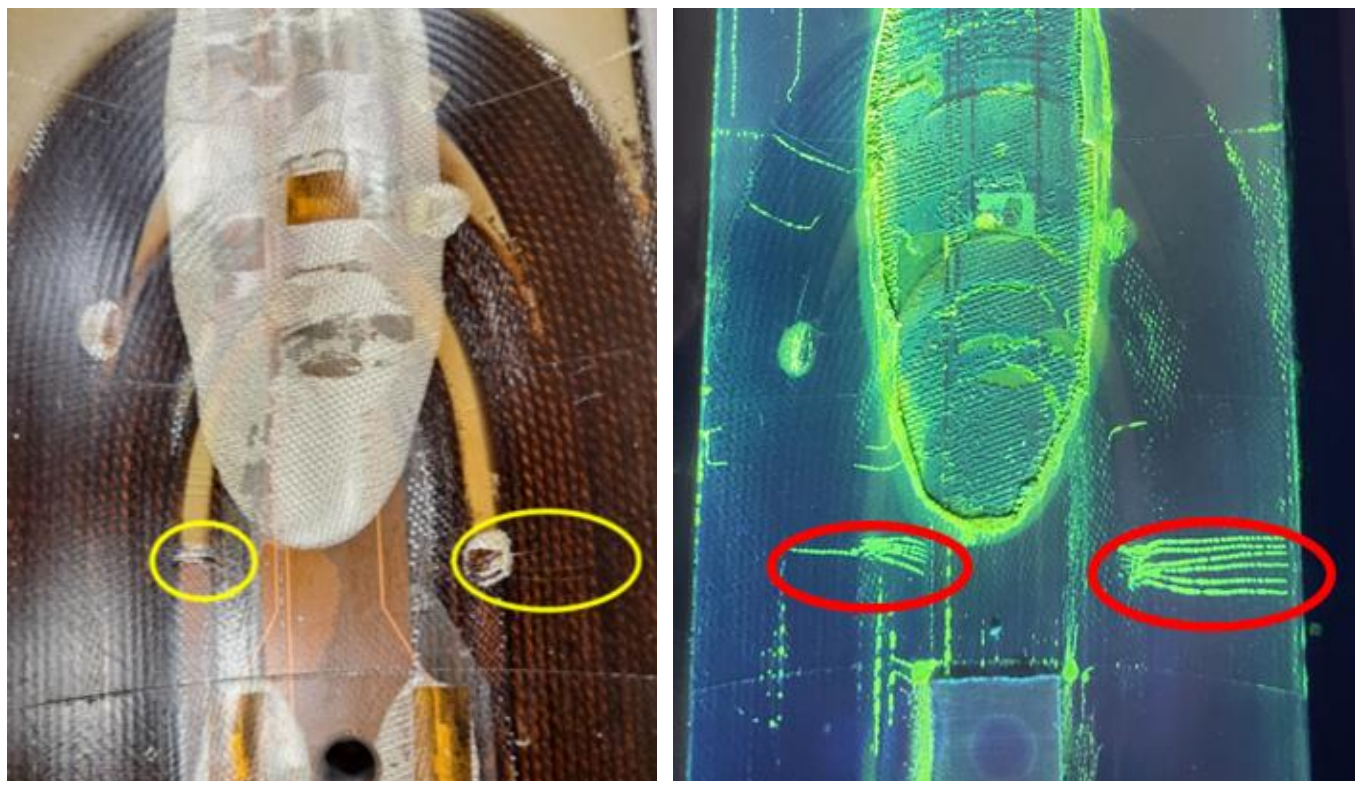

Fig. 8. Left: detail of coil 214 lead end. The yellow circles show bubble/delamination at each transition between wedge and end-spacer. Right: results of the dye penetrant test showing in the red circles cracks starting at the interface between wedges and end-spacers.

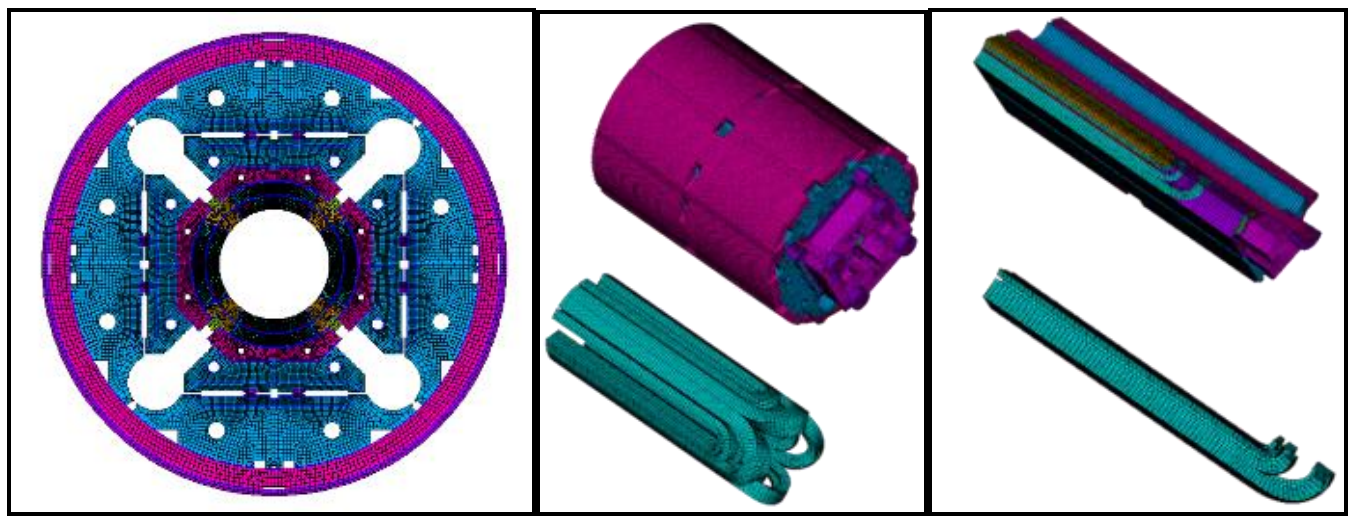

Fig. 9. Left: 360-degree, full cross-section 2D finite element model. Center: 360-degree, full cross-section 3D model. Right: octant 3D model.

## V. Finite Element Analysis

The effect on coil stress of closed pole-key gaps in Q3 was studied using three different finite element (FE) models: a 360-degree, full cross-section 2D model (see Fig. 9, left), a 360-degree, full cross-section 3D model (see Fig. 9, center), and an octant 3D model (see Fig. 9, right). Both 3D simulations reproduce an MQXF short model (MQXFS, with a 1.2 m magnetic length) to reduce the simulation time and increase the mesh density. The results in the end region were found to be consistent with those from the full-length model.

The results of the three analyses, presented in [22], are summarized in the following sub-sections.

### *A. 2D Finite Element Model*

According to the 2D FE model (which analyzes the magnet's straight section), no concerning stress singularity is observed in any of the quadrants. Nonetheless, the azimuthal pre-stress is significantly lower (~30 MPa reduction) in the quadrant with a closed pole-key gap (Q3) as compared to the adjacent quadrant (Q2 and Q4). In the opposite quadrant (Q1), the pre-stress is slightly lower than in Q2-Q4.

### *B. 3D Finite Element Model (360 degrees)*

The 3D FE model (360 degrees) confirms that the quadrant with closed pole-key gaps has a lower pre-stress than the quadrants with open pole-key gaps, which agrees with the 2D model predictions.

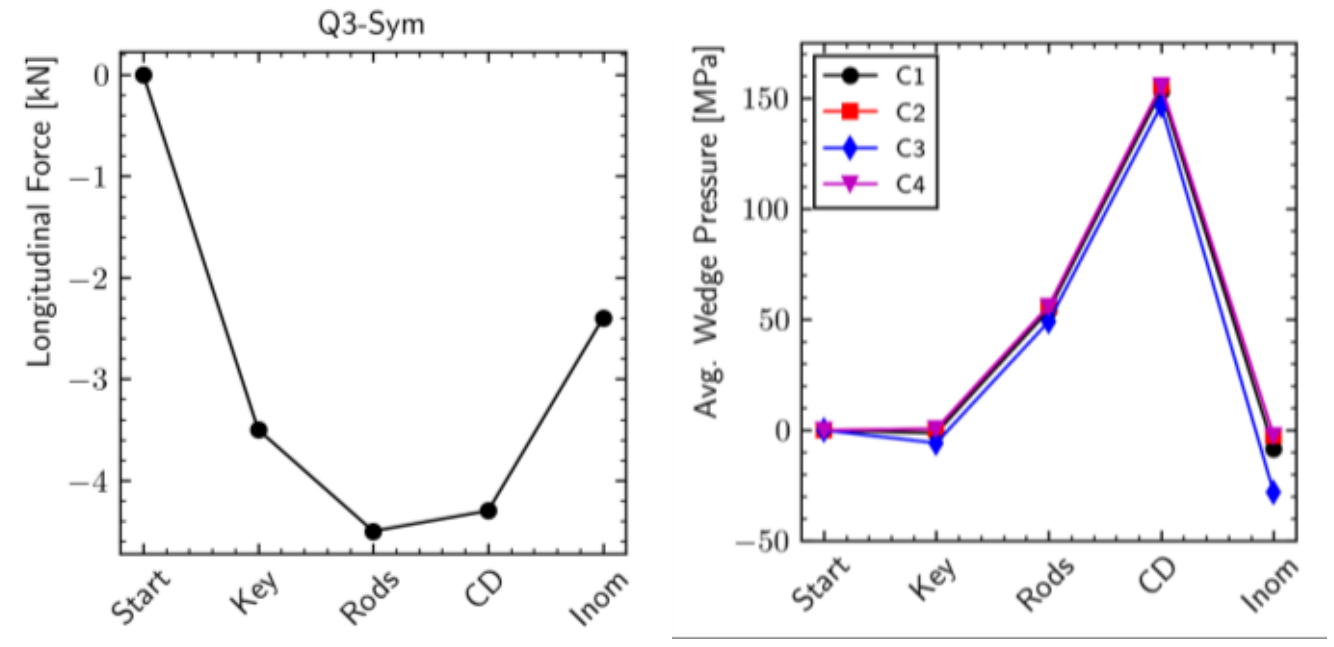

Fig. 10. Left: Difference between the force provided by the axial rods to the Q3 coil with respect to the case with an open pole-key gap. Right: average contact pressure (negative in tension) between wedge and end-spacer in the bonded case in the four quadrants.

In addition to the azimuthal stress in the straight section, the 3D model allows an investigation of the behavior of the four quadrants in the end region in the A07-A08 scenarios. If we focus on the axial pre-load system, the 3D model predicts that the axial force delivered by the end plate to the coil in Q3 is smaller as compared to the other quadrants (Fig. 10, left, shows the difference). This phenomenon can be explained as follows: the lower azimuthal pre-load in Q3 results in a smaller elongation caused by the Poisson effect, and in lower friction between the coil and the structure. Therefore, when compressed by the end plate, the Q3 coil contracts axially more than the other coils, which are better constrained axially by the higher frictional contact with the support structure. As a result, the coil Q3, being axially less rigid than the other three coils, receives less axial force from the end plate.

If now we look at the contact area between the wedge and the end-spacer in the coil's inner layer (indicated by the yellow circles of Fig. 8, left), we see that, if we assume that all of the coil's surfaces are bonded, a tension develops in Q3 during excitation, as shown in Fig. 10, right. More precisely, when Lorentz forces are applied, the average tension between wedge and end-spacer is about 30 MPa in quadrant 3, while it remains close to zero in the other quadrants.

In summary, a 30 MPa tension between wedge and end-spacer is observed in Q3 due to the lower azimuthal and axial pre-load caused by the closed pole-key gaps. This tension may cause cracking of the fiberglass-loaded epoxy between the wedge and the end-spacer, or separation between the epoxy and a metallic surface.

### *C. 3D Finite Element Model (Octant)*

The impact of a possible separation between the wedge and the end-spacer on the coil strain was studied using an octant model, which allows a more refined mesh as compared to the 360-degree model. Assuming that the there is no bonding between wedge and end-spacer (i.e. epoxy that has been cracked or separated from metal), a small gap opens and induces a local axial strain spike in the coil, which can exceed 0.4% (see Fig. 11). This strain level is considered sufficient to introduce permanent degradation due to filament failure in $Nb_3Sn$ strands [23]. The strain increase is larger in the pole block as compared to the mid-plane block.

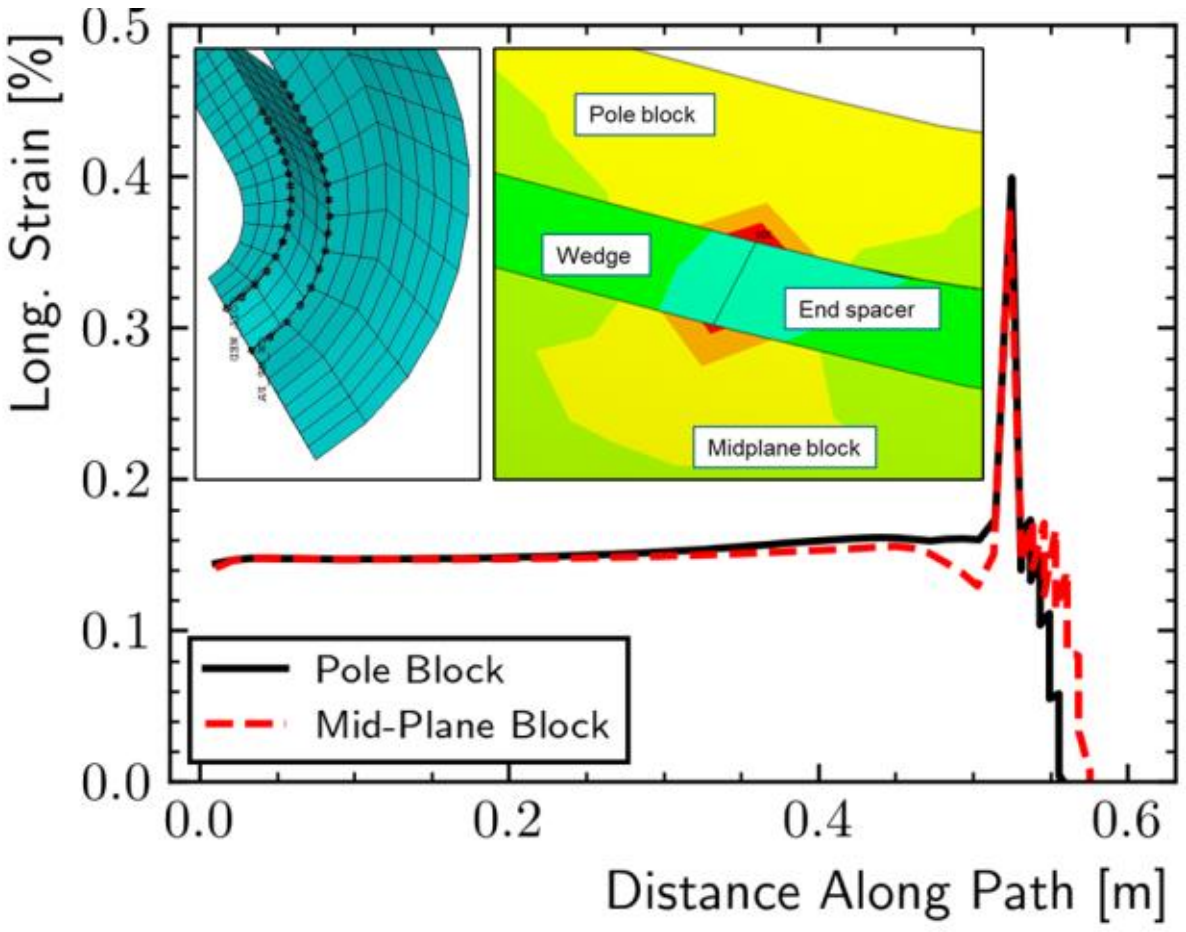

Fig. 11. Longitudinal strain if there is no bonding between wedge and end-spacer in the coil's inner layer lead end with asymmetric azimuthal preload. A short model was used for this analysis and the paths go from the magnet's center to its end.

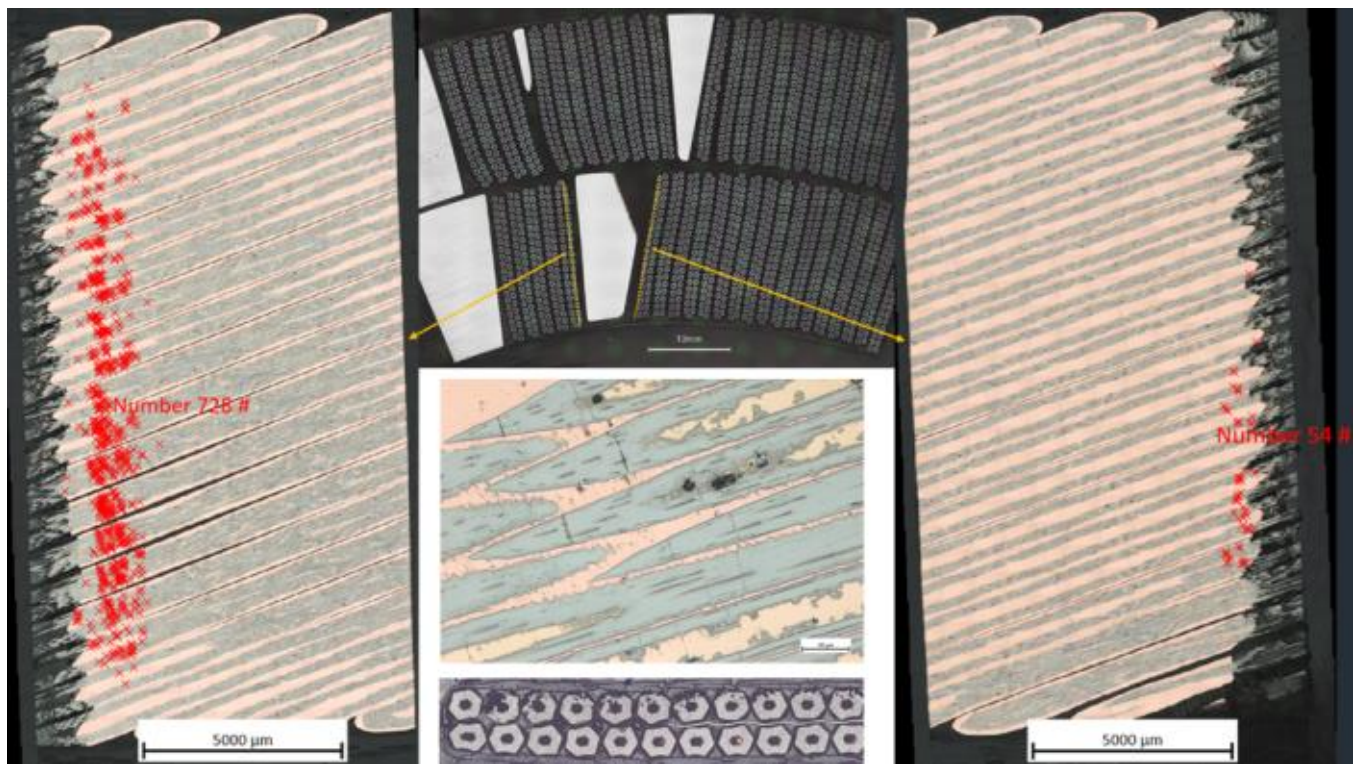

Fig. 12. Metallurgical inspection of cables adjacent to the wedge-spacer transition in the lead end non-layer-jump side of coil 214. Each red marker shows a cracked $Nb_3Sn$ filament (sub-element), as shown in the central and bottom pictures. Figure from [24].

In summary, the results of the FE analysis carried out with both 2D and 3D models can be summarized as follows: 1) closed pole-key gaps in quadrant 3 determine both lower azimuthal and axial pre-load as compared to the other quadrants; 2) as a result, high tension develops in the wedge/end-spacer interface, with a risk of epoxy cracking or de-bonding; and 3) in the case of a de-bonded wedge/end-spacer interface, the resulting gap creates a peak of axial strain in the coil turns adjacent to the wedge.

## VI. Metallurgical Inspection of Limiting Coil

As a result of all the observations and computations described in the previous sections, it was decided to perform micrographic analysis of the regions of the inner layer of the coil 214 lead end corresponding to the wedge to end-spacer transition on both coil sides. This analysis was performed at CERN and was part of a larger investigation described in [24]. In fact, the FE analysis described in Section V has shown the possibility of strain concentration in this region due to the unusual elongation of coil 214 caused by the closed gaps. Both visual inspection and die penetrant tests presented in Fig. 8 showed signs of strain concentration in these regions. Therefore, the cables most affected by this mechanism seemed to be those on both sides of each wedge to end-spacer transition.

In order to assess strand integrity along the whole wedge to end-spacer transition, it was decided to perform longitudinal cuts on the two cables facing the wedge to end-spacer transition on both sides (see Fig. 12, center top). The longitudinal cuts were to be performed along the cable side facing the transition (yellow lines). The results of the metallurgical inspection are presented in Fig. 12 in the left and right pictures where every red marker indicates a crack in an $Nb_3Sn$ filament (sub-element). The cracks are predominantly localized in the cable of the pole-block side, whereas no or fewer cracks were found in the cable included in the midplane block. This is consistent with the simulation results (see Fig. 11), which show a larger strain increase on the pole-block side. The central and bottom images show a detailed image of some cracks and the resulting collapsed filaments. In order to assess whether there is damage also on the second layer of strands (i.e., on the cable side opposite to the wedge-spacer transition), the same metallurgical analysis was performed on the second layer of strands, and not a single cracked filament was found on that cable sample.

A more focused investigation on the position of the cracks with respect to the wedge-spacer transition indicated that filament cracks are clearly clustered around the transition between the wedge and the epoxy resin (filled with S-2 glass). This is consistent with the observation in [24] that “the ceramic coating of the end-spacer provides good cohesion with the resin” and that the bonding between the copper wedge and the resin is not as strong.

TABLE I
Summary of the Metallurgical Inspection: # of Cracked Filaments

| Samples adjacent to the W-S transition from coil 214 LE | |
|---|---|
| 1) Layer-jump side, cable in midplane block, side adjacent to W-S transition | 0 |
| 2) Layer-jump side, cable in pole block, side adjacent to W-S transition | 532 |
| 3) Non-layer-jump side, cable in midplane block, side adjacent to W-S transition | 54 |
| 4) Non-layer-jump side, cable in pole block, side adjacent to W-S transition | 728 |
| 5) Same cable as in sample 4, side opposite to the W-S transition | 0 |

Table I presents a summary of this metallurgical inspection of cables, extracted from the coil 214 lead end adjacent to the wedge-spacer (W-S) transitions [24]. These results show that the strain field was concentered at the W-S transition, that the strain was significantly larger on the pole-block side than on the midplane-block side, that it affected only some cables adjacent to the W-S transition, and that it did not propagate to the second layer of strands (i.e., those further from the W-S transition). Similar results were obtained by an analysis of coil 214 that limited MQXFA08 [24].

## VII. Covid Impact

MQXFA07 and MQXFA08 were assembled at LBNL during the COVID pandemic. COVID prevention requirements caused two changes to MQXFA assembly procedures: 1) the coil-pack assembly procedure was changed in order to minimize the amount of time when two technicians were separated by less than six feet; and 2) beginning with MQXFA06, the technician who had been leading the coil-pack assembly operations through magnet MQXFA05 was removed from that task because of vaccination status.

The COVID lockdown caused a halt to the fabrication of several MQXFA coils, which had to be stored in unusual conditions. Coil 214, the MQXFA07 limiting coil, spent 14 weeks stored on the winding mandrel after its inner layer had been wound and cured. Usually, the outer layer is wound and cured a few days after the inner layer. Then, the tensioners keeping the coil stretched are removed. During the 14 weeks when coil 214’s inner layer was in storage with tensioners on, the bonding between the wedges and end-spacers may have degraded.

After the COVID lockdown fewer people than usual were allowed to access BNL and LBNL and requirements were put in place to minimize close contact. These factors reduced supervision during the completion of the fabrication of coils 213 and 214 and during the assembly of MQXFA07 and MQXFA08. The reduced supervision and other COVID requirements (i.e. distancing and mask requirements) may have been the occasion for undetected anomalies.

The changes described in the following section were introduced to assure successful magnet assembly even under strict COVID restrictions.

## VIII. Design and Specification Changes and their Implementation

As previously described, two important lessons were learned through MQXFA07/08 disassembly and FE analysis.

1) an asymmetry among pole-key gaps after coil-pack assembly is not be fixed by magnet preloading;
2) if the pole-key gaps are closed in a coil and open in the other three coils, the ends of the first coil are less preloaded, and tensile strain may develop in the turns close to the wedge-spacer transition.

In order to prevent this issue from occurring, the pole-key gaps were increased from 0.4 to 0.8 mm at each gap [25]. In addition, a specification was set for the minimum pole-key gap (0.6 mm) in each coil at each location [25]. MQXFA assembly procedures were changed to meet the tighter specifications [26].

By the time MQXFA07 was disassembled, the assembly of MQXFA10 was already in progress. Therefore, it was not possible to change the pole keys, and MQXFA10 implemented only the specification for the minimum pole-key gap that was set at 0.3 mm for this magnet. MQXFA11 implemented both changes, and all of the subsequent assemblies of MQXFA magnets will do likewise. The successful test of MQXFA10 and MQXFA11 (Fig. 1) demonstrated the effectiveness of these changes.

## IX. MQXFA Resilience

The truck transporting MQXFA11 from LBNL to BNL was involved in a highway crash [27]. During the crash, which lasted 1.5 s, MQXFA11 crate was displaced by about 2 m and the magnet shipping frame experienced several shocks in all directions. Most shocks were at or below 5 g, which is the maximum acceleration allowed by the MQXFA Handling and Shipping Requirements [28]. The largest shock was in vertical direction, causing a vibration that lasted 15 ms and reached 6 or 10 g as recorded by different devices (Piezo and DC-MEMS) on the same accelerometer unit [29]. The successful test of MQXFA11 demonstrated the resilience of MQXFA magnets.

## Conclusions

The AUP project has assembled ten MQXFA magnets. Eight have been tested in a vertical cryostat and two showed degradation during testing, with reverse temperature and ramp-rate dependence. A thorough investigation by AUP and CERN teams found the “smoking gun” and have understood the degradation mechanism. The main causes were: 1) assembly specifications focusing on average pole-key gaps without providing for a minimum for each gap, and 2) changes to assembly procedures due to COVID mitigation requirements. Based on lessons learned during the investigation, the pole-key gaps were increased, and the assembly specifications revised to set a minimum for each pole-key gap. The successful tests of MQXFA10 and MQXFA11 confirmed the effectiveness of these changes.

An endurance test performed on MQXFA05 showed that if an MQXFA magnet meets acceptance criteria, it can withstand a large number of quenches and several thermal cycles without experiencing any issues.

MQXFA11 was involved in a severe truck accident and withstood shocks up to 6-10 g. Its successful test has demonstrated the resilience of these magnets.

The MQXFA magnets are demonstrating that series production of $Nb_3Sn$ accelerator magnets is possible despite the strain sensitivity and brittleness of this conductor. Nonetheless, the following lesson learned should be considered: “*design and specifications for series production of $Nb_3Sn$ accelerator magnets must assure that all points in the acceptable-tolerance space are safe.*” In connection with this lesson, the analysis for assessing the acceptable-tolerance space should include the tolerances of all parts, sub-assemblies, and final assembly.